%% file: article_4_1.tex
\documentclass[
10pt, 
a4paper, 
oneside, 
headinclude,footinclude, 
BCOR5mm, 
]{scrartcl}

\input{structure.tex} 

\title{{Ineffectiveness of Dictionary Coding to Infer Predictability Limits of Human Mobility}} 

\author{\spacedlowsmallcaps{Yunheng Han\textsuperscript{1} Weiwei Sun\textsuperscript{1} Baihua Zheng\textsuperscript{2}}} 

\date{} 

\begin{document}


\renewcommand{\sectionmark}[1]{\markright{\spacedlowsmallcaps{#1}}} 
\lehead{\mbox{\llap{\small\thepage\kern1em\color{halfgray} \vline}\color{halfgray}\hspace{0.5em}\rightmark\hfil}} 

\pagestyle{scrheadings} 


\maketitle 

\setcounter{tocdepth}{2} 

\tableofcontents 




\section*{Abstract} 

Recently, a series of models have been proposed to predict future movements of people.
Meanwhile, dictionary coding algorithms are used to estimate the predictability limit of human mobility.
Although dictionary coding is optimal, it takes long time to converge.
Consequently, it is ineffective to infer predictability through dictionary coding algorithms.
In this report, we illustrate this ineffectiveness on the basis of human movements in urban space.



\let\thefootnote\relax\footnotetext{\textsuperscript{1} \textit{Fudan University, Shanghai, China}}
\let\thefootnote\relax\footnotetext{\textsuperscript{2} \textit{Singapore Management University, Singapore}}

\newpage 


\section{Introduction}

Understanding and predicting human mobility is of great importance for a wide spectrum of services and applications, e.g., transportation planning, emergency management and various location-based services. Many of the related works focus on building different models to predict human mobility and the ultimate goal is to improve the accuracy of the prediction models. This leads to a fundamental measurement: \emph{predictability}, which quantifies the degree to which human mobility is predictable. For example, a piece of very well-received work~\cite{limits-science} estimates the predictability based on dictionary coding, and it claims that the prediction accuracy of the next location of a certain mobile phone user is at most $93\%$, no matter what prediction approach is used. In this report, we present a counterexample to that highly-cited study and illustrate that the predictability estimated based on dictionary coding is not accurate. 

To be more specific, we study the predictability of human mobility in urban road networks. The advances in GPS-enabled mobile devices and pervasive computing techniques have generated massive trajectory data. For instance, Qiangsheng Taxi Company Limited, one of the largest taxi companies in Shanghai, obtained more than $3\times 10^{10}$ trajectory records in April, 2015~\cite{soda}. The large amount of trajectory data provide opportunities to further enhance urban transportation systems. 

For example, with the help of the large collection of trajectory data and the availability of digital maps, we could adopt data-driven approach to model map-matched trajectories. Given a road network that captures the roads in a modern city and a trajectory recording the movement of a moving object in the road network, trajectory modeling is to model the likelihood of a given trajectory that passes $k$ edges sequentially. As to be detailed in Section~\ref{sec:review}, trajectory modeling tries to compute the transition probability $P(e_{i+1}|e_1, e_2, \cdots, e_i)$, the probability that a driver who drives from edge $e_1$ to $e_i$ via edges $e_2$, $e_3$, $\cdots$ takes the edge $e_{i+1}$. As presented in~\cite{modeling-ijcai}, Recurrent Neural Networks (RNN) based models are able to predict the transition probability with $87.8\%$ accuracy. 

On the other hand, we adopt dictionary coding in this paper as an alternative to model trajectory and propose multiple novel techniques to improve the coding performance. Based on the experimental study result obtained from the real dataset used by~\cite{modeling-ijcai}, the predictability limit of dictionary coding based approach is around $82\%$ which is lower than the accuracy of $87.8\%$ achieved by RNN based models proposed in~\cite{modeling-ijcai}. It demonstrates the ineffectiveness and limitations of dictionary coding in predicting human mobility in road networks. Meanwhile, it also provides evidence for illustrating the potential mistakes in the previous dictionary coding based study on predictability limit of human mobility~\cite{limits-science}.

The rest of the report is organized as follows.
First, we review related work in Section~\ref{sec:review}.
Then, we describe several techniques to improve the coding performance in Section~\ref{sec:methods}.
Finally, we present and discuss the experiments in Section~\ref{sec:results} and conclude the results in Section~\ref{sec:conclusion}.

\section{Literature Review}
\label{sec:review}

In this section, we review the work on three topics closely relevant to this report, including \emph{trajectory models in road networks}, \emph{trajectory compression}, and \emph{predictability limits of human mobility}.

We first begin with some preliminaries about trajectories and road networks.
A trajectory is the path that a moving object follows in space, as a function of time.
In practice, raw trajectory data contains series of sample points, where each sample point contains a position $p_i$ and the corresponding time stamp $t_i$, as presented in Definition~\ref{def:raw-traj}. 

\begin{definition}[Trajectory] 
	\label{def:raw-traj}
	A trajectory $T$ is a sequence of $|T|$ tuples in the form of $\langle p_{1},t_{1}\rangle$, $\langle p_{2},t_{2}\rangle$, $\cdots$, $\langle p_{|T|},t_{|T|}\rangle$, where $p_{i}$ describes the position of the moving object in terms of longitude and latitude at the time stamp $t_i$.
\end{definition} 

A road network is modeled as a directed graph $G(V, E)$,
where crossroads and road segments are represented by vertices and edges respectively.
Note that vehicles in urban spaces are restrained by the underlying road networks, which differ from objects that move arbitrarily in common spaces.
Notwithstanding, raw trajectory data is often biased from the network because of sampling errors.
Consequently, map-matching algorithms should be applied to align the sample points with the road network.
The results of the map-matching process are sequences of edges in $G$.  In the rest of this report, we use map-matched trajectory and edge sequence interchangeably as they both refer to a sequence of edges in the road network passed by a moving object along a trajectory. 

\begin{definition}[Road network] 
	A road network is a directed graph $G(V, E)$, in which $V$ is the set of vertices and $E$ is the set of edges.
\end{definition} 

\begin{definition}[Map-matched trajectory] 
	A map-matched trajectory is a sequence of $n$ road segments, i.e., a series of edges in $G$, in the form of $e_1$, $e_2$,$\cdots$,$e_n$.
\end{definition}

\subsection{Trajectory Modeling}

In trajectory models, it is essential to figure out the probability of a given trajectory.
Unfortunately, the information entropy of trajectories tends to be infinity as the precision of data grows~\cite{compress-tods}. 
In other words, the uncertainty of sample points is high.
As a result, the probability distribution of original trajectories, in the form of $\langle p_{i},t_{i}\rangle$ sequence as defined in Definition~\ref{def:raw-traj}, can hardly be determined.

That being said, however, it is possible to compute the probability of a map-matched trajectory $T$~\cite{modeling-ijcai}, i.e.,
\begin{equation}
	P(T)=P(e_1)\prod_{i=1}^{n-1}P(e_{i+1}|e_1,e_2,\cdots,e_i),
\end{equation}
because map-matched trajectories are integer sequences, whose information entropy is limited.
Such map-matched trajectory models have been adopted to solve many problems relevant to location-based services~\cite{recovery,osogami,xue,yuan}. 
Two models based on Recurrent Neural Networks are designed in~\cite{modeling-ijcai}, to compute the probability of a given edge sequence.
Other approaches~\cite{zhengandni,Ziebart} use first-order Markov chains to model map-matched trajectories, which have been proved to be insufficient~\cite{Srivatsa}. 

Computing the probability of an edge sequence is equivalent to predicting road transition, i.e., computing the probability $P(e_{i+1}|e_1,e_2,\cdots,e_i)$.
To the best of our knowledge, the RNN-based approaches~\cite{modeling-ijcai} are the most effective, in terms of prediction accuracy.
Nevertheless, all approaches mentioned above only give lower bounds of the prediction accuracy.
The upper bounds, i.e., limits of predictability, can be bounded by information entropy~\cite{shannon}, which is introduced in Section~\ref{subsec:limit}.

\subsection{Trajectory Compression}

Data compression is to represent data with fewer bits, so that the storage cost is reduced.
During the past decades, a lot of trajectory data compression techniques~\cite{compress-tods, press-vldb, icde-cinct} have been proposed to meet the requirements of trajectory data storage.
Here, we only review some recent works on processing map-matched trajectory data.

PRESS~\cite{press-vldb} is the first piece of work on compressing map-matched trajectories.
It implements \emph{Shortest Path Compression} and \emph{Frequent Path Compression} to reduce the size of edge sequences. \emph{Minimal Entropy Labeling (MEL)} proposed in~\cite{compress-tods} is the first method to convert original edge sequences into label sequences, which significantly reduces the alphabet size from the original $|E|$ to the max out-degree $D$ of any vertex with $D\ll |E|$. 
MEL allocates a unique label to each out edge of every vertex in the road network, according to frequency of edges in the data set.
This labeling method is optimal for the symbol-by-symbol coding algorithms like Huffman coding, because it is able to minimize the entropy of label sequences.

Different from MEL, \emph{Relative Movement Labeling (RML)} proposed in~\cite{icde-cinct} takes into account the dependence between two consecutive labels.
This is achieved by constructing a directed graph called ET-graph, in which a vertex $v_{ET}$ represents an edge $e_i$ in the data set, while the succeeding vertices $u_{ET}$ of $v_{ET}$ represent preceding edges $e_{i-1}$ in edge sequences.
Afterwards, a label is allocated to each edge $\langle v_{ET}, u_{ET} \rangle$ in the ET-graph, according to frequency of bi-grams $\langle e_{i-1},e_i\rangle$ in the data set.

If we still apply symbol-by-symbol coding after labeling, RML does not outperform MEL.
However, if arithmetic coding is applied, coding length after RML will be smaller, because arithmetic coding may encode several symbols at once. 
Furthermore, Lempel-Ziv algorithms converge faster after RML than after MEL, though the eventual coding length remains the same since the entropy does not change during the labeling processes.

There is a close connection between prediction and data compression. 
Optimal prediction is equivalent to optimal data compression.
On the one hand, if one can predict probabilities of the next symbol, on the basis of the historical data, then optimal coding length is achieved by applying arithmetic coding.
On the other hand, given an optimal compressor, one can use the symbol that compresses the best for the prediction.
Thereby, data compression is used as a benchmark for ``general intelligence''.

\subsection{Predictability of Human Mobility}
\label{subsec:limit}

The first method to address the predictability of human mobility studies location data of mobile phone users~\cite{limits-science}.
The data set contains trajectories of $5\times10^4$ users, in which the length of each trajectory is truncated to $2352$.

The method used in~\cite{limits-science} is based on information theory and coding algorithms. Information entropy is a measure of uncertainty~\cite{shannon, cover}.
The lower the entropy is, the lower the uncertainty is, and thereby the higher the predictability is.
Note that locations of mobile phone users are matched to positions of routing towers, so the number of locations and thus the entropy are limited, which makes it possible to predict future movements of users.

The authors first present the relationship between predictability and information entropy. Given a history trajectory $h_{n-1}=\{x_1,x_2,\cdots,x_{n-1}\}$ that denotes the sequence of towers at which user $i$ was observed at each consecutive hourly interval, the best guess of the next location is to choose the location $x_n$ with the highest probability, so the optimal prediction accuracy based on $h_{n-1}$ is 
\begin{equation}
\label{eq:pii}
\pi(h_{n-1}) = sup_{x_n}\{P(X_n=x_n|h_{n-1})\}.
\end{equation}
Considering all the history trajectories of length $n-1$, the optimal prediction accuracy of the $n$-th location is 
\begin{equation}
\label{eq:pin}
\Pi(n)=\sum_{h_{n-1}}P(h_{n-1})\pi(h_{n-1}).
\end{equation}
As the length of a historical trajectory $h_n$ tends to be infinity, the overall limit of prediction accuracy is derived in Equation~(\ref{eq:pi}). 
\begin{equation}
\label{eq:pi}
\Pi=\lim_{n \to \infty}\frac{1}{n}\sum_{i=1}^{n}\Pi(i).
\end{equation}
It is proven that the limit of predictability can be bounded by the entropy rate:
\begin{equation}
\label{ieq}
H(\mathcal{X}) \leq H(\Pi) + (1-\Pi)\log_2(S-1),
\end{equation}
where $H(\mathcal{X})$ is the entropy rate, $H(p) = -p\log_2p - (1-p)\log_2(1-p)$ is the binary entropy, and $S$ is the total number of the routing towers, i.e., the maximum number of different locations visited by the user.

Next, the authors use Lempel-Ziv algorithms~\cite{lempel,LZW} to estimate the entropy rate $H(\mathcal{X})$.
According to optimality of Lempel-Ziv coding~\cite{ziv}, its average coding length per symbol is asymptotically no greater than the entropy rate.
Therefore, average coding length is an approximation of the entropy rate, as the length of trajectories tends to be infinity.

However, the experimental studies presented in~\cite{compress-tods} imply that this method has some drawbacks.
On the one hand, the speed that average coding length converges to the entropy rate is very slow.
One the other hand, average coding length cannot be less than the entropy, which is the limit of data compression~\cite{shannon}. 
Thereby, the estimated entropy rate is usually positively biased, especially when $S$ is large or trajectories are short.
If the estimated entropy rate is greater than its true value, the value of $\Pi$ will be negatively biased.
Consequently, the upper bound of prediction accuracy estimated by the coding method is erroneous.

There are other works applying the coding method to determine the predictability of traffic conditions~\cite{limits-traffic} and vehicle staying time~\cite{limits-staying}. 
The two problems are similar because they both study attributes of certain areas over time, such as the average speed and the staying time.


\section{Methodology}
\label{sec:methods}


As it is mentioned in Section~\ref{sec:review}, the reason that average coding length of Lempel-Ziv algorithms cannot converge to the entropy rate is two-fold, i.e., the alphabet size is large and the input trajectories are short. For example, a road network of a modern city has tens or hundreds of thousands edges,  and a trajectory in our dataset passes by 35 edges on average.  
Consequently, transformation methods that can transform trajectories into other representations are necessary before applying dictionary code. 
In the following, we propose two steps, namely \emph{alphabet reduction} and \emph{sequence construction}, as solutions to tackle the two issues stated previously. 


Let $h_{n-1}$ denote the history map-matched trajectory that passes ($n-1$) edges, and the optimal prediction accuracy is 
\begin{equation}
\label{equ:opt_tra_pred}
\pi(h_{n-1}) = sup_{e}\{P(e_{n} = e|h_{n-1})\}. 
\end{equation}
Note Equation~(\ref{equ:opt_tra_pred}) is consistent with Equation~(\ref{eq:pii}), although each tower $x_i$ appearing in $h_{n-1}$ is replaced by an edge $e_i$ passed by the map-matched trajectory.
In other words, the definitions of the predictability $\Pi(n)$ and $\Pi$ in the context of map-matched trajectories are similar to those in~\cite{limits-science}, i.e., in Equation~(\ref{eq:pin}) and Equation~(\ref{eq:pi}).




\subsection{Alphabet Reduction}

The alphabet size of inputs, i.e., the number of edges in $G$, is so large that the coding algorithm can hardly find any repeated patterns.
Hence Lempel-Ziv algorithms fail to divide inputs into long distinct phrases.
Consequently, the compression will be ineffective, i.e., the average coding length will not be a good approximation of the entropy rate.

In order to address the problem resulting from the large alphabet, we plan to extend the labeling methods proposed in~\cite{compress-tods,icde-cinct}.
There is no information loss after labeling process, because all label sequences can be converted to the original edge sequences, and vice versa.
In other words, there is an injective function between edge sequences and label sequences, so labeling processes do not change the entropy.
Nevertheless, the alphabet size is reduced and thus the data becomes easier to compress after labeling processes.
Therefore, the coding algorithms converge faster on label sequences than on edge sequences.

As it is mentioned in Section~\ref{sec:review}, MEL considers labels to be independently distributed, while RML only takes into account dependence between two consecutive labels.
The labeling method can be further improved by considering dependence among more consecutive labels, so that the coding algorithms will be more effective.
More precisely, we can label edges according to frequency of longer sub-sequences in the data set.
However, the sub-sequences should not be too long, otherwise their frequency will be too low because of data sparsity, conversely leading to ineffective coding.

\subsection{Sequence Construction}

The second problem is that trajectories are too short.
Although Lempel-Ziv algorithms are optimal, it takes long time for average coding length to converge to the entropy.
In this project, every map-matched trajectory represents a single trip of a person via a car.
Therefore, different from the time series studied in~\cite{limits-traffic, limits-staying}, whose average length grows as data sets become larger, the average length of trips is relatively fixed, even though the data collection is huge.
For example, the average length of map-matched trajectories in our data set is about $35$, while the data set contains more than $7\times10^7$ trips.
As a result, such short inputs cannot guarantee convergence of the coding length.

In order to address this issue, our proposal is to construct a long sequence based on the trajectory data set, and then take the long sequence as the input of the coding algorithm, so that the average coding length is able to converge to the entropy.
Given a set of $k$ trajectories $T_i$s with each $T_i = \{e_{i,1}, e_{i,2},\cdots,e_{i,n_i}\}$, we can convert it into a single long sequence by connecting all trajectories together, i.e., $T_{input}= \{T_1,0,T_2,0,\cdots,T_k,0\}$, where $0$ represents the end of a trajectory.
Note that the entropy of the long sequence will not be smaller than that of real trajectories since this construction does not cause any information loss. 
However, the long sequence is not a real trajectory, so the entropy will be different.
In other words, the entropy of the long sequence must be greater than the entropy of real trajectories, which will result in erroneous estimation.

We plan to design an approach to construct an arbitrarily long sequence, and the entropy of its distribution is theoretically no greater than the entropy of real trajectories.
Thereby, the coding length is able to converge and the estimation of the predictability limit will be a valid upper bound.
In addition, the entropy of the construction should not be too negatively biased, otherwise the estimated upper bound will be meaningless.
Furthermore, the construction approach should be compatible with alphabet reduction, so that both of the problems will be solved.

In most data compression and indexing methods, trajectory data are stored in the form of $T_1$,$0$,$T_2$,$0$,$\cdots$,$T_k$,$0$, so that there is no information loss.
However, this format results in an intangible entropy value.
On the other hand, the entropy rate can be calculated directly, if we connect real trajectories directly without using any extra symbols, i.e., in the form of $T_1,T_2,\cdots,T_k$.


In the following, we generate a long sequence by picking up trajectories randomly from the data set, after which we build the relationship between entropy of the constructed sequence and  that of the real trajectory data.
First, we sample the value of trajectory length $n_i$ (i.e., $|T_i|$) from $N$, the length distribution of trajectory data.
Second, we sample a trajectory whose length is exactly $n_i$ from the conditional distribution $T|N=n_i$.
The sampling strategy is equivalent to sample trajectories directly from $T$.

Let $T_i = e_{i,1}, e_{i,2},\cdots,e_{i,n_i}$ denote the $i$-th trajectory in the constructed sequence.
Given any sample sequence of trajectory length, i.e., $n_1$, $n_2$, $\cdots$, $n_i$, the entropy rate is written as
\begin{align*}
H(\mathcal{T}) &= \lim_{k \to \infty} \frac{1}{k}\sum_{k'=1}^{k}H(e_{i',j'}|e_{1,1}, e_{1,2},\cdots,e_{1,n},\cdots,e_{i',1}, e_{i',2},\cdots,e_{i',j'-1})
\end{align*}
and here we have $k = \sum_{i'=1}^{i-1}n_{i'} + j$ and $k' =  \sum_{i''=1}^{i'-1}n_{i''} + j'$, because trajectories vary in length.
Afterwards, the entropy rate is calculated as:
\begin{align*}
H(\mathcal{T}) &= \lim_{k \to \infty} \frac{1}{k}(\sum_{i'=1}^{i-1}\sum_{j'=1}^{n_{i'}}H(e_{i',j'}|e_{i',1}, \cdots,e_{i',j'-1})+\sum_{j'=1}^{j}H(e_{i,j'}|e_{i,1}, \cdots,e_{i,j'-1}))\\
&=\lim_{k \to \infty} \frac{1}{k}(\sum_{i'=1}^{i-1}H(e_{i',1},e_{i',2} \cdots,e_{i',n_{i'}})+H(e_{i,1},e_{i,2},\cdots,e_{i,j}))\\
&=\lim_{k \to \infty} \frac{1}{\sum_{i'=1}^{i-1}n_{i'} + j}(\sum_{i'=1}^{i-1}H(T|N=n_{i'})+H(e_{i,1}, \cdots,e_{i,j}|N=n_i))
\end{align*}
As $j \leq n_i$ and $H(e_{i,1}, \cdots,e_{i,j}|N=n_i)\leq H(T|N=n_i)$ for any $j$, the limit does not change if we remove them from the dominator and the numerator respectively:
\begin{align*}
H(\mathcal{T}) &= \lim_{k \to \infty}
 \frac{\sum_{i'=1}^{i-1}H(T|N=n_{i'})}{\sum_{i'=1}^{i-1}n_{i'}}=\lim_{i \to \infty}
 \frac{\sum_{i'=1}^{i}H(T|N=n_{i'})}{\sum_{i'=1}^{i}n_{i'}}
\end{align*}
The strong law of large numbers implies that the sample average converges almost surely to the expected value:
\begin{align*}
Pr[\lim_{i \to \infty} \frac{\sum_{i'=1}^{i}H(N)}{i} = E[H(N)] &=H(T|N)] = 1\\
Pr[\lim_{i \to \infty} \frac{\sum_{i'=1}^{i}N}{i}& = E[N]] = 1
\end{align*}
where it is noteworthy that the entropy $H(N)$ is a random variable, and $Pr[H(N)=H(T|N=n_i)]=Pr[N=n_i]$.

According to properties of limits, we have 
\begin{align*}
Pr[\lim_{i \to \infty} \frac{\sum_{i'=1}^{i}H(N)}{\sum_{i'=1}^{i}N}& = \frac{H(T|N)}{E[N]}] = 1
\end{align*}
In other words, the entropy rate and thus the average coding length of Lempel-Ziv coding is $H(T|N)/E[N]$:
\begin{align*}
H(\mathcal{T}) &= \lim_{i \to \infty}
\frac{\sum_{i'=1}^{i}H(T|N=n_{i'})}{\sum_{i'=1}^{i}n_{i'}} = \frac{H(T|N)}{E[N]}
\end{align*}

The predictability of trajectory data whose length is limited can be also bounded by the entropy, by extending the results in~\cite{limits-science}.
Fano's inequality shows that 
\begin{align*}
H(e_i|e_1,e_2,\cdots,e_{i-1})&\leq H_F(\Pi(i))\\
\frac{1}{n}\sum_{i=1}^{n}H(e_i|e_1,e_2,\cdots,e_{i-1})&\leq \frac{1}{n}\sum_{i=1}^{n}H_F(\Pi(i))\\
\frac{H(T|N=n)}{n}&\leq \frac{1}{n}\sum_{i=1}^{n}H_F(\Pi(i))
\end{align*}
According to Jensen's inequality and the fact that $H_F(p) = H(p)+(1-p)\times \log_2(S-1)$ is a convex function, the average predictability is bounded by the entropy:
\begin{align}
\frac{H(T|N=n)}{n}&\leq H_F(\frac{1}{n}\sum_{i=1}^{n}\Pi(i))\\
\sum_{n}Pr[N=n]\frac{H(T|N=n)}{n}&\leq \sum_{n}Pr[N=n]H_F(\frac{1}{n}\sum_{i=1}^{n}\Pi(i))\\
\label{ieq2}
E[\frac{H(N)}{N}]&\leq H_F(\sum_{n}(\frac{Pr[N=n]}{n}\sum_{i=1}^{n}\Pi(i)) =H_F(\Pi)
\end{align}
However, $E[H(N)/N]$ is not bounded by $E[H(N)]/E[N]$.
Consequently, it is not proper to directly connect all trajectories of various lengths to construct the long sequence.

Now we consider the case in which all the trajectories have the same length $n$.
Then the long sequence will be $e_{1,1}$, $e_{1,2}$,$\cdots$,$e_{1,n}$, $e_{2,1}$, $e_{2,2}$,$\cdots$,$e_{2,n}$, $\cdots$, $e_{i,1}$, $e_{i,2}$,$\cdots$,$e_{i,n}$, $\cdots$.
After construction, we can easily separate the long sequence into original trajectories,
even though the long sequence does not include extra symbols like $0$, 
because each trajectory is of length $n$. 
As a result, there is no information loss after construction, and the entropy will be no smaller than the real entropy.
More precisely, the entropy rate of the long sequence is
\begin{align*}
H(\mathcal{T}) &= \lim_{k \to \infty} \frac{1}{k}H(e_{1,1}, e_{1,2},\cdots,e_{1,n},\cdots,e_{i,1}, e_{i,2},\cdots,e_{i,j})\\
&= \lim_{k \to \infty} \frac{1}{k}\sum_{k'=1}^{k}H(e_{i',j'}|e_{1,1}, e_{1,2},\cdots,e_{1,n},\cdots,e_{i',1}, e_{i',2},\cdots,e_{i',j'-1})
\end{align*}
where $k = n\times (i-1) + j$ and $k' = n\times (i'-1) + j'$.
If we pick up trajectories from the data set randomly during the construction, i.e., the order of real trajectories in the long sequence is random, there will be less relevance between each part of the long sequence.
In other words, we sample trajectories $T_i$ from the identical distribution $T$ independently.
On the basis of this assumption, the entropy rate will be
\begin{align*}
H(\mathcal{T}) &= \lim_{k \to \infty} \frac{1}{k}\sum_{k'=1}^{k}H(e_{i',j'}|e_{1,1}, e_{1,2},\cdots,e_{1,n},\cdots,e_{i',1}, e_{i',2},\cdots,e_{i',j'-1})\\
&= \lim_{k \to \infty} \frac{1}{k}(\sum_{i'=1}^{i-1}\sum_{j'=1}^{n}H(e_{i',j'}|e_{i',1}, \cdots,e_{i',j'-1})+\sum_{j'=1}^{j}H(e_{i,j'}|e_{i,1}, \cdots,e_{i,j'-1}))\\
&=\lim_{k \to \infty} \frac{1}{k}(\sum_{i'=1}^{i-1}H(e_{i',1}, e_{i',2},\cdots,e_{i',n})+H(e_{i,1}, e_{i,2},\cdots,e_{i,j}))\\
&=\lim_{k \to \infty} \frac{(i-1)\times H(T)}{n\times (i-1) + j}+\lim_{k \to \infty}\frac{H(e_{i,1}, e_{i,2},\cdots,e_{i,j})}{n\times (i-1) + j}\\
&=\lim_{i \to \infty} \frac{(i-1)\times H(T)}{n\times (i-1)}+0 =\frac{H(T)}{n}
\end{align*}
The entropy rate and hence the average coding length will be $H(T)/n$, if we only use trajectories of length $n$ to construct the long sequence.


Now we sum up the overall approach to the problem.
First, we categorize all trajectories in the data set according to their length, so that they will form many groups.
Second, we construct long sequences with respect to each group and apply Lempel-Ziv coding and determine the value $H(T|N=n)/n$ for each $n$.
Third, we estimate $E[H(N)/N]$ as well as $\Pi$ according to Inequation~(\ref{ieq2}).

\subsection{Fusion of Strategies}

Sequence construction is compatible with the labeling strategy.
We illustrate the idea by a simple proof.
According to the fact that 
\begin{align*}
\frac{H(e_1,e_2,\cdots,e_n)}{n}\leq H_F(\frac{1}{n}\sum_{i=1}^{n}\Pi(i))
\end{align*}
and
\begin{align*}
H(e_1,e_2,\cdots,e_n)=H(e_1,l_1,\cdots,l_{n-1}),
\end{align*}
we have
\begin{align*}
\frac{H(e_1,l_1,\cdots,l_{n-1})}{n}\leq H_F(\frac{1}{n}\sum_{i=1}^{n}\Pi(i)).
\end{align*}
Thereby,  
\begin{align*}
\frac{H(l_1,\cdots,l_{n-1})}{n} \leq H_F(\frac{1}{n}\sum_{i=1}^{n}\Pi(i)).
\end{align*}
$\hat{H}_n=H(l_1,\cdots,l_{n-1})/(n-1)$ can be estimated through sequence construction, so we have
\begin{align*}
\frac{n-1}{n}\cdot \hat{H}_n \leq H_F(\frac{1}{n}\sum_{i=1}^{n}\Pi(i)) = H_F(\Pi_n)
\end{align*}
Note that the prediction accuracy of the first edge is nearly $0$, and we only care about the predictability of the following edges, so
\begin{align*}
\hat{\Pi}_n = \frac{1}{n-1}\sum_{i=2}^{n}\Pi(i) \approx \frac{1}{n-1}\sum_{i=1}^{n}\Pi(i) = \frac{n-1}{n}\cdot \Pi_n
\end{align*}
Finally, the predictability limit is
\begin{align*}
\Pi = \sum_n Pr[N=n]\cdot \hat{\Pi}_n 
\end{align*}

\section{Experiments}
\label{sec:results}

In this section, we present the experimental results, to prove the ineffectiveness of the approach proposed in~\cite{limits-science}.
We first introduce the data set used in the experimental study.
The data set consists of real trajectory data from Qiangsheng Taxi, which contains more than $7\times10^7$ trips.
The road network of Shanghai is extracted from OpenStreetMap, which contains $60,200$ edges and $28,620$ vertices.
This dataset is exactly the same as the dataset used in~\cite{modeling-ijcai}, and the prediction accuracy achieved by the models proposed in~\cite{modeling-ijcai} is 87.8\%. 
All the algorithms are implemented with C programming language and run on a computer with Intel Core i$7$-$6820$HK CPU@$3.30$ GHz and $32$ GB memory.


Then we apply the Lempel-Ziv-Welch algorithm~\cite{LZW}, one of the most commonly used Lempel-Ziv algorithms, directly on the map-matched trajectory data.
The average coding length per symbol is $15.94$ bits.
Note that it only takes $\lceil\log_2 |E|\rceil$, i.e., $16$ bits to represent an edge, if we use na\"ive sequential coding.
Consequently, the LZW algorithm can hardly compress the trajectory data, let alone does the average coding length converge to the entropy rate.
This demonstrates the poor performance of the optimal coding on map-matched trajectory data.

As explained in Section~\ref{sec:methods}, we reduce the alphabet size through the labeling methods.
The average coding length drops to $1.15$ bits if MEL is applied, while that becomes $1.12$ bits if RML is applied, with both being much smaller than that of direct Lempel-Ziv coding.
However, the coding length is still not satisfactory, because it leads to the estimated upper bounds of $81.9\%$ and $82.3\%$ respectively, according to Inequation~\ref{ieq}.

Moreover, we apply both sequence construction and alphabet reduction.
As the alphabet size is merely $6$ and the constructed sequences all contain at least $10^4$ labels ($5\times10^4$ on average), the performance of dictionary coding is significantly improved.
Nevertheless, the estimated upper bound is $85.3\%$, which is still lower than $87.8\%$, the best prediction accuracy achieved by trajectory models so far. 


\section{Conclusion}
\label{sec:conclusion}

To draw a conclusion, the coding length of Lempel-Ziv algorithms does not converge to the entropy rate of trajectories, though the two transformation strategies are applied.
Consequently, it is ineffective to infer predictability limits through the approach based on dictionary coding.
In other words, the approach proposed in~\cite{limits-science} is problematic.
In the near future, we plan to design a new approach to infer predictability limits.


\renewcommand{\refname}{\spacedlowsmallcaps{References}} 

\bibliographystyle{unsrt}

\bibliography{sample.bib} 


\end{document}

%% file: structure.tex
\usepackage[
nochapters, 
beramono, 
eulermath,
pdfspacing, 
dottedtoc 
]{classicthesis} 

\usepackage{arsclassica} 

\usepackage[T1]{fontenc} 

\usepackage[utf8]{inputenc} 

\usepackage{graphicx} 
\graphicspath{{Figures/}} 

\usepackage{enumitem} 

\usepackage{lipsum} 

\usepackage{subfig} 

\usepackage{amsmath,amssymb,amsthm} 

\usepackage{varioref} 

\usepackage{comment}

\theoremstyle{definition} 
\newtheorem{definition}{Definition}

\theoremstyle{plain} 

\theoremstyle{remark} 

\hypersetup{
colorlinks=true, breaklinks=true, bookmarks=true,bookmarksnumbered,
urlcolor=webbrown, linkcolor=RoyalBlue, citecolor=webgreen, 
pdftitle={}, 
pdfauthor={\textcopyright}, 
pdfsubject={}, 
pdfkeywords={}, 
pdfcreator={pdfLaTeX}, 
pdfproducer={LaTeX with hyperref and ClassicThesis} 
}